\documentclass[12pt,eqno,epsf]{article}
\usepackage{amsmath,amssymb,graphicx,slashbox}
\numberwithin{equation}{section}

\def\ignore#1{{}}

\newcounter{sxn}

\newcounter{axn}

\date{}

\newdimen\mybaselineskip
\renewcommand{\thefootnote}{\arabic{footnote}}

\newcommand{\beeq}{\begin{equation}}
\newcommand{\eneq}{\end{equation}}
\newcommand{\beqn}{\begin{eqnarray}}
\newcommand{\eeqn}{\end{eqnarray}}

\newcommand{\alp}{\alpha}
\newcommand{\bt}{\beta}
\newcommand{\gm}{\gamma}

\newcommand{\dlt}{\delta}

\newcommand{\vep}{\varepsilon}
\newcommand{\tht}{\theta_{\rm H}}
\newcommand{\vth}{\vartheta}

\newcommand{\lmd}{\lambda}
\newcommand{\Lmd}{\Lambda}
\newcommand{\sgm}{\sigma}

\newcommand{\vph}{\varphi}

\newcommand{\Omg}{\Omega}
\newcommand{\sth}{s_\theta}
\newcommand{\cth}{c_\theta}

\newcommand{\be}{\begin{equation}}
\newcommand{\ee}{\end{equation}}
\newcommand{\bea}{\begin{eqnarray}}
\newcommand{\eea}{\end{eqnarray}}
\newcommand{\eql}{\!\!\!&=\!\!\!&}

\newcommand{\defa}{\!\!\!&\equiv\!\!\!&}

\newcommand{\exch}{\leftrightarrow}
\newcommand{\simgt}{\stackrel{>}{{}_\sim}}
\newcommand{\simlt}{\stackrel{<}{{}_\sim}}

\newcommand{\tl}[1]{\tilde{#1}}
\newcommand{\bdm}[1]{{\mbox{\boldmath $#1$}}}
\newcommand{\tr}{{\rm tr}}

\newcommand{\diag}{{\rm diag}}
\newcommand{\der}{\partial}
\newcommand{\dr}{\!\!d}

\newcommand{\ie}{{\it i.e.}}
\newcommand{\id}{\mbox{\boldmath $1$}}

\newcommand{\brkt}[1]{\left( #1 \right)}
\newcommand{\brc}[1]{\left\{ #1 \right\}}
\newcommand{\sbk}[1]{\left[ #1 \right]}
\newcommand{\abs}[1]{\left| #1 \right|}

\newcommand{\cA}{{\cal A}}
\newcommand{\cB}{{\cal B}}

\newcommand{\cD}{{\cal D}}

\newcommand{\cL}{{\cal L}}
\newcommand{\cM}{{\cal M}}

\newcommand{\cO}{{\cal O}}
\newcommand{\cP}{{\cal P}}

\newcommand{\cW}{{\cal W}}

\newcommand{\mKK}{m_{\rm KK}}

\newcommand{\NP}[1]{{\it Nucl.~Phys.}~{\bf #1}}
\newcommand{\PL}[1]{{\it Phys.~Lett.}~{\bf #1}}

\newcommand{\MPL}[1]{{\it Mod.~Phys.~Lett.}~{\bf #1}}

\newcommand{\PR}[1]{{\it Phys.~Rev.}~{\bf #1}}
\newcommand{\PRL}[1]{{\it Phys.~Rev.~Lett.}~{\bf #1}}
\newcommand{\PTP}[1]{{\it Prog.~Theor.~Phys.}~{\bf #1}}

\newcommand{\JH}[1]{{\it JHEP}~{\bf #1}}

\begin{document}
\thispagestyle{empty}

\baselineskip=12pt

{\small \noindent 
\hfill OU-HET 627/2009}

{\small \noindent \hfill  RIKEN-TH-153}

\baselineskip=35pt plus 1pt minus 1pt

\vskip 2.0cm

\begin{center}
{\Large \bf Weak boson scattering in Gauge-Higgs Unification}\\

\vspace{2.0cm}
\baselineskip=20pt plus 1pt minus 1pt

\normalsize

{\bf Naoyuki\ Haba},${}^1\!${\def\thefootnote{\fnsymbol{footnote}}
\footnote[1]{\tt e-mail address: haba@het.phys.sci.osaka-u.ac.jp}} 
{\bf Yutaka\ Sakamura}${}^2\!${\def\thefootnote{\fnsymbol{footnote}}
\footnote[2]{\tt e-mail address: sakamura@riken.jp}}
{\bf and
Toshifumi\ Yamashita}${}^3\!${\def\thefootnote{\fnsymbol{footnote}}
\footnote[3]{\tt e-mail address: yamasita@eken.phys.nagoya-u.ac.jp}}

\vspace{.3cm}
${}^1${\small \it Department of Physics, Osaka University, 
Toyonaka, Osaka 560-0043, Japan} \\
${}^2${\small \it RIKEN, Wako, Saitama 351-0198, Japan} \\
${}^3${\small \it Department of Physics, Nagoya University, 
Nagoya 464-8602, Japan}
\end{center}

\vskip 2.0cm
\baselineskip=20pt plus 1pt minus 1pt

\begin{abstract}
The scattering amplitude for the longitudinal weak bosons is investigated 
in the $SU(3)$ gauge-Higgs unification as a function of the scattering energy, 
the Wilson line phase~$\tht$ and the warp factor. 
The $\tht$-dependence of the amplitude is quite different in the flat 
and the warped spacetimes. 
Generically the amplitude is enhanced for $\tht=\cO(1)$ 
in the warped case while it is almost independent of $\tht$ in the flat case. 
This indicates the tree-level unitarity is violated in the warped case 
at a lower scale than that in the flat case. 
\end{abstract}


\newpage
\section{Introduction}
The origin of the electroweak symmetry breaking is still a mystery 
since the Higgs boson has not been discovered yet. 
Extra dimensions open up new possibility for it. 
For example, it can occur by nontrivial boundary conditions 
or the Wilson line phases along the extra dimensions, 
such as the Higgsless models~\cite{Higgsless} 
or the gauge-Higgs unification models~\cite{GHU1}-\cite{GHU5}. 
Since these models are based on higher dimensional gauge theories 
and are thus nonrenormalizable, 
they should be interpreted as effective theories of more fundamental theories, 
which are valid below certain cutoff energy scales. 
This implies that the tree-level unitarity is violated 
at some scale, which is identified with the cutoff scale of the model. 
When we work in a higher dimensional theory, we have to know 
the cutoff scale of the theory in order to ensure the validity 
of the perturbative calculation. 

The tree-level unitarity is usually discussed by evaluating 
the scattering amplitudes of the longitudinally polarized 
weak bosons~$W^\pm_L$ and $Z_L$ at tree-level. 
In the standard model, the Higgs boson plays an important role 
for the recovery of the tree-level unitarity. 
If it is sufficiently heavy and decoupled, 
the scattering amplitudes for $W^\pm_L$ and $Z_L$ grow as $E^2$ where $E$ is 
the energy scale of the scattering, 
and exceed the unitarity bound at some scale around 1~TeV. 
This means that 
the perturbative calculation is no longer reliable above the scale. 
In the Higgsless models, the tree-level unitarity is recovered by 
the Kaluza-Klein (KK) modes of the gauge bosons~\cite{Higgsless} 
instead of the Higgs boson in the standard model, 
and the unitarity violation delays up to $\cO(10~{\rm TeV})$ 
when the compactification scale is assumed to be around 1~TeV. 

The situation is more complicated in the gauge-Higgs unification models 
because they have the Higgs mode 
(the fluctuation of the Wilson line phase~$\tht$) as well as the KK modes 
of the gauge bosons, both of which participate 
in the unitarization of the theory. 
In these models, both the coupling constants and the KK mass scale  
depend on $\tht$ and thus the scattering amplitudes for $W^\pm_L$ and $Z_L$ 
have nontrivial $\tht$-dependences. 
Especially the gauge-Higgs unification 
in the Randall-Sundrum warped spacetime~\cite{warpGHU}-\cite{GHU:HS2} 
is interesting because the $WWH$ and the $ZZH$ couplings 
($H$ stands for the Higgs mode) 
deviate from the standard model values 
and vanish for some specific values of $\tht$, such as $\pi$ or $\pi/2$, 
depending on the models~\cite{GHU:HS1,GHU:HS2}. 
For such values of $\tht$, the Higgs mode cannot participate in 
the unitarization of the weak boson scattering. 
Therefore it is important to understand 
the $\tht$-dependence of the scattering amplitudes for $W^\pm_L$ and $Z_L$ 
in order to estimate the cutoff scale of the models from the violation 
of the tree-level unitarity. 
This issue is discussed in Ref.~\cite{FPR} and 
some qualitative behaviors of the amplitude are clarified.  

In this paper, we investigate various behaviors of the 
scattering amplitude more quantitatively
by numerical calculations. 
We focus on the process:~$W^+_L+W^-_L\to Z_L+Z_L$ 
in the gauge-Higgs unification model based on the five-dimensional (5D) 
$SU(3)$ gauge theory on $S^1/Z_2$ as the simplest example. 
Although this model gives the wrong value of the Weinberg angle~$\theta_W$
and thus is not realistic, it has a lot of common features 
among the gauge-Higgs unification models. 
Hence it is a good starting point to understand the behaviors of 
the amplitudes peculiar to the gauge-Higgs unification. 
The Wilson line phase~$\tht$, which corresponds to 
the vacuum expectation value (VEV) of the Higgs field 
in the standard model, is dynamically determined 
at one-loop order if the whole matter content of the model is specified. 
In the following discussion, we do not specify the fermion sector 
and treat $\tht$ as a free parameter 
because we are interested in the tree-level amplitude. 

The paper is organized as follows. 
In Sec.~\ref{model}, we briefly review the $SU(3)$ gauge-Higgs unification model 
and provide necessary ingredients to calculate the scattering amplitude 
for the weak bosons. 
In Sec.~\ref{scattering}, we provide explicit expressions of 
the scattering amplitudes for the longitudinal weak bosons and 
for the (would-be) Nambu-Goldstone (NG) bosons, 
and show their behaviors as functions of 
$E$, $\tht$ and the warp factor. 
Sec.~\ref{summary} is devoted to the summary and discussions. 
In Appendix~\ref{basis_fcn}, we give definitions and explicit forms 
of the basis functions used in the text. 
In Appendix~\ref{5Dpropagator}, we derive the 5D propagators 
of the gauge fields.

\section{$\bdm{SU(3)}$ model} \label{model}
\subsection{Set-up}
We consider the 5D $SU(3)$ gauge theory compactified on $S^1/Z_2$ 
as the simplest example of the gauge-Higgs unification. 
Arbitrary background metric with 4D Poincar\'{e} symmetry can be written as 
\be
 ds^2=G_{MN}dx^Mdx^N=e^{-2\sgm(y)}\eta_{\mu\nu}dx^\mu dx^\nu+dy^2, 
\ee
where $M,N=0,1,2,3,4$ are 5D indices and $\eta_{\mu\nu}=\diag(-1,1,1,1)$. 
The fundamental region of $S^1/Z_2$ is $0\leq y\leq L$. 
The function~$e^{\sgm(y)}$ is a warp factor, which is normalized as 
$\sgm(0)=0$. 
For example, $\sgm(y)=0$ in the flat spacetime, 
and $\sgm(y)=ky$ ($0\leq y\leq L$) in the Randall-Sundrum spacetime~\cite{RS}, 
where $k$ is the inverse AdS curvature radius. 

The 5D gauge field~$A_M$ is decomposed as 
\be
 A_M=\sum_{\alp=1}^8A_M^\alp\frac{\lmd^\alp}{2}, 
\ee
where $\lmd^\alp$ are the Gell-Mann matrices. 
The 5D Lagrangian is 
\bea
 \cL \eql \sqrt{-G}\sbk{-\tr\brc{\frac{1}{2}G^{ML}G^{NP}F_{MN}F_{LP}
 +\frac{1}{\xi}\brkt{f_{\rm gf}}^2}}+\cdots, 
 \label{5D_Lagrangian}
\eea
where $\sqrt{-G}\equiv\sqrt{-\det(G_{MN})}=e^{-4\sgm}$, 
$F_{MN}\equiv\der_M A_N-\der_N A_M-ig_5\sbk{A_M,A_N}$ 
($g_5$ is the 5D gauge coupling), and $\xi$ is a dimensionless parameter.  
The ellipsis denotes the ghost and the matter sectors, which are 
irrelevant to the following discussion. 
The gauge-fixing function~$f_{\rm gf}$ is chosen as 
\bea
 f_{\rm gf} \eql e^{2\sgm}\brc{\eta^{\mu\nu}\der_\mu A_\nu
 +\xi \cD_y^c\brkt{e^{-2\sgm}A_y}}, \nonumber\\
 \cD_y^c A_M \defa \der_y A_M-ig_5\sbk{A_y^{\rm bg},A_M},
\eea
where $A_y^{\rm bg}(y)$ is the classical background of $A_y(x,y)$. 

The boundary conditions for the gauge field is written as 
\bea
 A_\mu(x,-y) \eql P_0A_\mu(x,y)P_0^{-1}, \;\;\;
 A_\mu(x,L+y) = P_LA_\mu(x,L-y)P_L^{-1}, \nonumber\\
 A_y(x,-y) \eql -P_0A_y(x,y)P_0^{-1}, \;\;\;
 A_y(x,L+y) = -P_LA_y(x,L-y)P_L^{-1}, 
\eea
where $P_0$ and $P_L$ are unitary matrices satisfying 
the relation~$P_0^2=P_L^2=1$. 
By choosing them as $P_0=P_L=\diag(-1,-1,1)$, 
the $Z_2$-parity eigenvalues~$(P_0,P_L)$ of the gauge fields become
\bea
 A_\mu \eql \begin{pmatrix} (+,+) & (+,+) & (-,-) \\
 (+,+) & (+,+) & (-,-) \\ (-,-) & (-,-) & (+,+) \end{pmatrix}, \;\;\;\;\;
 A_y = \begin{pmatrix} (-,-) & (-,-) & (+,+) \\
 (-,-) & (-,-) & (+,+) \\ (+,+) & (+,+) & (-,-) \end{pmatrix}. 
\eea
Note that only $(+,+)$ fields can have zero-modes when perturbation theory 
is developed around the trivial configuration~$A_M=0$. 
Thus the $SU(3)$ gauge symmetry is broken to $SU(2)\times U(1)$ 
at tree-level. 
The zero-modes of $A_y$ form an $SU(2)$-doublet 
4D scalar~$(A_y^4+iA_y^5,A_y^6+iA_y^7)$, which plays a role of 
the Higgs doublet in the standard model whose VEV breaks 
$SU(2)\times U(1)$ to $U(1)_{\rm EM}$. 
They yield non-Abelian Aharonov-Bohm phases 
(Wilson line phases) when integrated along the fifth dimension. 
By using the residual $SU(2)\times U(1)$ symmetry, 
we can always push the nonvanishing VEV into $A_y^7$. 
Then the Wilson line phase~$\tht$ is given by 
\be
 \tht =g_5\int_0^L\dr y\;A_y^{\rm bg\,7}(y). 
\ee

\subsection{Mode expansion}
The mode expansion of the 5D gauge fields is performed 
in a conventional way (see Ref.~\cite{GHU:HNSS}, for example). 
For the following discussion, it is convenient to move to 
the momentum representation for the 4D part while remain the coordinate 
representation for the fifth dimension. 
Then the 5D gauge fields are expanded into the KK modes as 
\bea
 \tl{A}_\mu^\alp(p,y) \eql \sum_n u_n^\alp(y)A_\mu^{(n)}(p)
 +\sum_n w_n^\alp(y)p_\mu A_{\rm S}^{(n)}(p), \nonumber\\ 
 \tl{A}_y^\alp(p,y) \eql \sum_n v_n^\alp(y)\vph^{(n)}(p). 
\eea
Here we have moved to the Scherk-Schwarz basis, in which $\tl{A}_y^{\rm bg}=0$. 
It is related to the original basis by the gauge transformation, 
\be
 \tl{A}_M = \Omg A_M\Omg^{-1}-\frac{i}{g_5}\brkt{\der_M\Omg}\Omg^{-1}, 
 \label{gauge_trf}
\ee
with 
\be
 \Omg(y) \equiv \cP\exp\brc{-ig_5\int_0^y\dr y'\;A_y^{\rm bg\,7}(y')\cdot
 \frac{\lmd^7}{2}}. 
 \label{def:Omg}
\ee
The symbol~$\cP$ stands for the path ordered operator from left to right. 
Notice that $\tl{A}_\mu^\alp(p,y)$ are decomposed into two parts, 
according to their polarization. 
In the above expression, $A_\mu^{(n)}(p)$ are polarized as 
$p^\mu A_\mu^{(n)}(p)=0$ and include the transverse and 
the longitudinal modes, which are physical for the massive modes. 
On the other hand, $A_{\rm S}^{(n)}(p)$ are unphysical scalar modes. 
The gauge-scalar modes~$\vph^{(n)}(p)$ are also unphysical besides 
the zero-mode. 

The mode functions~$w_n^\alp(y)$ and $v_n^\alp(y)$ are related to each other by 
\bea
 v_n^\alp(y) \eql \frac{d}{dy}w_n^\alp(y), \nonumber\\
 w_n^\alp(y) \eql -\frac{\xi}{\tl{m}_n^2}
 \frac{d}{dy}\brc{e^{-2\sgm(y)}v_n^\alp(y)}, 
\eea
where $\tl{m}_n$ are common mass eigenvalues for $A_{\rm S}^{(n)}$ and 
$\vph^{(n)}$. 
These relations hold irrespective of the value of the gauge parameter~$\xi$. 
When $\xi=1$, they are further related to $u_n^\alp(y)$ as 
\be
 u_n^\alp(y) = m_nw_n^\alp(y), \;\;\;
 m_n = \tl{m}_n,  
\ee
where $m_n$ are mass eigenvalues for $A_\mu^{(n)}$. 

According to the transformation properties under the unbroken $U(1)_{\rm EM}$ 
symmetry and the rotation by a constant matrix~$\Omg(L)$, 
the gauge fields are classified into the charged 
sector~$(A_M^{1\pm i2},A_M^{4\pm i5})\equiv 
(A_M^1\pm iA_M^2,A_M^4\pm iA_M^5)/\sqrt{2}$ 
and the neutral sectors~$(A_M^{3'},A_M^6)$, $A_M^7$ and $A_M^{8'}$, where
\be
 \begin{pmatrix} A_M^{3'} \\ A_M^{8'} \end{pmatrix}
 \equiv \begin{pmatrix} -\frac{1}{2} & \frac{\sqrt{3}}{2} \\
 -\frac{\sqrt{3}}{2} & -\frac{1}{2} \end{pmatrix}
 \begin{pmatrix} A_M^3 \\ A_M^8 \end{pmatrix}. 
\ee
The $W$, $Z$ bosons and the photon are identified with the lowest modes 
in the $(A_\mu^{1\pm i2},A_\mu^{4\pm i5})$-,  
the $(A_\mu^{3'},A_\mu^6)$- and the $A_\mu^{8'}$-sectors, respectively. 

The mode functions for the $W$ boson are calculated as
\bea
 u_W^1(y) \defa u_0^{1+i2}(y) = N_W \cth S_0(L,m_W)C_0(y,m_W), 
 \nonumber\\
 u_W^4(y) \defa u_0^{4+i5}(y) = -N_W \sth C_0(L,m_W)S_0(y,m_W), 
\eea
where $\cth\equiv\cos(\tht/2)$, $\sth\equiv\sin(\tht/2)$, 
and $C_0(y,m)$, $S_0(y,m)$ are the basis functions defined in 
Appendix~\ref{basis_fcn}. 
The mode functions for the $Z$ boson are 
\bea
 u_Z^{3'}(y) \defa u_0^{3'}(y) = N_Z\cos\tht S_0(L,m_Z)C_0(y,m_Z), 
 \nonumber\\
 u_Z^6(y) \defa u_0^6(y) = -N_Z\sin\tht C_0(L,m_Z)S_0(y,m_Z). 
\eea
Here the normalization constants~$N_W$ and $N_Z$ are determined by 
\be
 \int_0^L\dr y\;\brc{(u_W^1)^2+(u_W^4)^2} = 1, \;\;\;\;\;
 \int_0^L\dr y\;\brc{(u_Z^{3'})^2+(u_Z^6)^2} = 1, 
\ee
and the $W$ and the $Z$ boson masses~$m_W$ and $m_Z$ are 
the lowest solutions of 
\bea
 \frac{1}{m_W}\brc{C'_0(L,m_W)S_0(L,m_W)
 +m_We^{\sgm(L)}\sth^2} \eql 0, \nonumber\\
 \frac{1}{m_Z}\brc{C'_0(L,m_Z)S_0(L,m_Z)
 +m_Ze^{\sgm(L)}\sin^2\tht} \eql 0, 
 \label{eq_for_spectrum}
\eea
respectively. 
The prime denotes the derivative for $y$. 
In the flat spacetime, for example, they are written for $0<\tht<\pi$ as 
\be
 m_W = \frac{\tht}{2L}, \;\;\;\;\;
 m_Z = \frac{\abs{\pi-\abs{\pi-2\tht}}}{2L}. \label{mWZ:flat}
\ee

For the corresponding gauge-scalar modes, 
the mode functions~$v_W^{1,4}(y)$, $v_Z^{3',6}(y)$ and their masses are 
obtained similarly.

\subsection{5D propagators} \label{expr:5D_prop}
For the purpose of calculating the scattering amplitude, 
it is convenient to use the 5D propagators defined in a mixed 
momentum/position representation~\cite{GP}. 
It describes the propagation of the entire KK tower of excitations 
carrying the 4D momentum~$p$ between two points~$y$ and $y'$ 
in the extra dimension. 
This approach has an advantage that we need not explicitly calculate 
the mass eigenvalues for the modes propagating in the internal lines 
of the Feynmann diagrams nor sum over contributions from 
the infinite number of KK modes.\footnote{
This approach is also useful for models with continuum spectra~\cite{Unhiggs}.
} 
The explicit forms of the 5D propagators can be obtained 
by using the formula~(\ref{expr:G_T}). 
For the charged sector~$(A_\mu^{1+i2},A_\mu^{4+i5})$, 
the 5D propagator is calculated as 
\bea
 G_{\rm T<}(y,y') \eql \frac{e^{2\sgm(L)}}{\det_{(1,4)}\cW}
 \begin{pmatrix} \abs{p}e^{\sgm(L)}C_0(y)S_0(L)C_L(y') & \\
 & -S_0(y)C'_0(L)S_L(y') \end{pmatrix} \nonumber\\
 &&-\frac{\sth\abs{p}e^{3\sgm(L)}}{\det_{(1,4)}\cW}
 \begin{pmatrix} \sth C_0(y)S_0(y') & \cth C_0(y)S_0(y') \\
 \cth S_0(y)C_0(y') & -\sth S_0(y)C_0(y') \end{pmatrix}, \nonumber\\
 G_{\rm T>}(y,y') \eql \brc{G_{\rm T<}(y',y)}^t, 
\eea
where $\abs{p}\equiv\sqrt{-p^2}$, and 
$\det_{(1,4)}$ is the determinant of $\cW$ defined in (\ref{def:cW}) 
restricted to the $(A_\mu^{1+i2},A_\mu^{4+i5})$-sector and   
\bea
 {\rm det}_{(1,4)}\cW \eql -\abs{p}e^{\sgm(L)}
 \brc{C'_0(L)S_0(L)+\abs{p}e^{\sgm(L)}\sth^2}. 
\eea
In the above expressions, we have omitted $\abs{p}$ from the arguments. 
For the $(A_\mu^{3'},A_\mu^6)$-sector, the propagator is obtained 
from the above expressions by replacing $\tht/2$ with $\tht$. 
The propagators in the other sectors are calculated as 
\bea
 G_{\rm T<}^{77}(y,y',\abs{p}) \eql 
 e^{\sgm(L)}\frac{S_0(y,\abs{p})S_L(y',\abs{p})}{\abs{p}S_0(L,\abs{p})}, 
 \nonumber\\
 G_{\rm T<}^{8'8'}(y,y',\abs{p}) \eql 
 -e^{2\sgm(L)}\frac{C_0(y,\abs{p})C_L(y',\abs{p})}{C'_0(L,\abs{p})}. 
\eea

\section{Weak boson scattering} \label{scattering}
Now we consider the scattering of the weak bosons. 
The two-body scattering amplitudes are functions of 
the total energy~$E$ and the scattering angle~$\phi$ 
in the center-of-mass frame. 
For the elastic scattering of the $W$ bosons, there is 
an infrared singularity at $\phi=0$ due to the $t$-channel diagram 
exchanging the massless photon. 
Although such a singularity will be cancelled when 
the soft-photon emissions and higher loop corrections 
are taken into account, it involves some 
technical difficulties to obtain a finite value of the amplitude 
for the forward scattering. 
In order to avoid such difficulties, we consider the scattering process:  
$W_L^+(p_1)+W_L^-(p_2)\to Z_L(p_3)+Z_L(p_4)$ in the following. 
The subscript~$L$ denotes the longitudinal polarization.

\subsection{Scattering amplitude}
As mentioned in Sec.~\ref{expr:5D_prop}, 
the scattering amplitudes are easily calculated by utilizing 
the 5D propagators. 
The scattering amplitude~$\cA$ is expressed by 
\be
 \cA = \cA^{\rm C}+\cA^{\rm V}+\cA^{\rm S},  \label{expr:cA}
\ee
where $\cA^{\rm C}$, $\cA^{\rm V}$ and $\cA^{\rm S}$ are contributions 
from the contact interaction, the exchange of the vector modes 
and that of the gauge-scalar modes, respectively. 
They are given by 
\bea
 \cA^{\rm C} \eql -\frac{ig_5^2}{4}\int_0^L\dr y\;
 \brc{(u_W^1)^2+(u_W^4)^2}
 \brc{(u_Z^{3'})^2+(u_Z^6)^2} \nonumber\\
 &&\times \brc{2(\vep_1\cdot\vep_2)(\vep_3^*\cdot\vep_4^*)
 -(\vep_1\cdot\vep_3^*)(\vep_2\cdot\vep_4^*)
 -(\vep_1\cdot\vep_4^*)(\vep_2\cdot\vep_3^*)}, \label{A^C} \\
 \cA^{\rm V} \eql ig_5^2\sum_{\alp,\bt=1,4}
 \int_0^L\dr y\int_0^L\dr y'\;
 U_{WZ}^\alp(y)G_{\rm T}^{\alp\bt}(y,y',\abs{p_{13}})U_{WZ}^\bt(y')
 P_{1324} \nonumber\\
 &&+ig_5^2\sum_{\alp,\bt=1,4}\int_0^L\dr y\int_0^L\dr y'\;
 U_{WZ}^\alp(y)G_{\rm T}^{\alp\bt}(y,y',\abs{p_{14}})U_{WZ}^\bt(y')
 P_{1423},  \label{A^V} \\
 \cA^{\rm S} \eql ig_5^2\int_0^L\dr y\;
 Y_{WW}^7(y)Y_{ZZ}^7(y)\frac{(\vep_1\cdot\vep_2)(\vep_3^*\cdot\vep_4^*)}
 {p_{12}^2}  \nonumber\\
 &&+ig_5^2\sum_{\alp=1,4}\int_0^L\dr y\;
 Y_{WZ}^\alp(y)Y_{WZ}^\alp(y)
 \brc{\frac{(\vep_1\cdot\vep_3^*)(\vep_2\cdot\vep_4^*)}{p_{13}^2} 
 +\frac{(\vep_1\cdot\vep_4^*)(\vep_2\cdot\vep_3^*)}{p_{14}^2}}, 
 \label{A^S}
\eea
where $\vep_i$ ($i=1,2,3,4$) are the polarization vectors, 
$p_{12}\equiv p_1+p_2$, $p_{13}\equiv p_1-p_3$, $p_{14}\equiv p_1-p_4$, 
\bea
 P_{1324} \defa \brc{2(p_1\cdot\vep_3^*)\vep_1+2(p_3\cdot\vep_1)\vep_3^*
 -(\vep_1\cdot\vep_3^*)(p_1+p_3)}^\mu
 \brkt{\eta_{\mu\nu}-\frac{p_{13\mu}p_{13\nu}}{p_{13}^2}} \nonumber\\
 &&\times\brc{2(p_2\cdot\vep_4^*)\vep_2+2(p_4\cdot\vep_2)\vep_4^*
 -(\vep_2\cdot\vep_4^*)(p_2+p_4)}^\nu, \nonumber\\
 P_{1423} \defa \brc{2(p_1\cdot\vep_4^*)\vep_1+2(p_4\cdot\vep_1)\vep_4^*
 -(\vep_1\cdot\vep_4^*)(p_1+p_4)}^\mu
 \brkt{\eta_{\mu\nu}-\frac{p_{14\mu}p_{14\nu}}{p_{14}^2}} \nonumber\\
 &&\times\brc{2(p_2\cdot\vep_3^*)\vep_2+2(p_3\cdot\vep_2)\vep_3^*
 -(\vep_2\cdot\vep_3^*)(p_2+p_3)}^\nu, 
\eea
and the functions in the integrands are defined as 
\bea
 U_{WZ}^1 \eql \frac{1}{2}\brkt{u_W^1 u_Z^{3'}+u_W^4 u_Z^6}, \;\;\;\;\;
 U_{WZ}^4 = \frac{1}{2}\brkt{u_W^1 u_Z^6-u_W^4 u_Z^{3'}}, \nonumber\\
 Y_{WW}^7 \eql e^{-2\sgm}\brc{(u_W^1)'u_W^4-u_W^1(u_W^4)'}, \;\;\;\;\;
 Y_{ZZ}^7 = 2e^{-2\sgm}
 \brc{(u_Z^{3'})'u_Z^6-u_Z^{3'}(u_Z^6)'}, \nonumber\\
 Y_{WZ}^1 \eql \frac{e^{-2\sgm}}{2}
 \brc{(u_W^1)'u_Z^{3'}-u_W^1(u_Z^{3'})'
 +(u_W^4)'u_Z^6-u_W^4(u_Z^6)'}, \nonumber\\
 Y_{WZ}^4 \eql \frac{e^{-2\sgm}}{2}\brc{(u_W^1)'u_Z^6-u_W^1(u_Z^6)'
 -(u_W^4)'u_Z^{3'}+u_W^4(u_Z^{3'})'}.  
\eea
Here we have used the relation~$p_i\cdot\vep_i(p_i)=0$ ($i=1,2,3,4$). 

The first and the second lines in (\ref{A^V}) correspond to 
the $t$-channel and the $u$-channel diagrams exchanging the $W$ boson 
and its KK modes, respectively.  
The above expression of the amplitude is a result of 
a cancellation between the gauge-dependent part~$G_{\rm S}(y,y',\abs{p})$ 
in the propagator of the vector modes and 
the gauge-scalar propagator~$G_{yy}(y,y',\abs{p})$. 
This cancellation occurs due to the relation~(\ref{rel:Gyy-Gs}) 
and makes the resultant amplitude gauge-independent. 
The contribution~$\cA^{\rm S}$ is a remnant of the cancellation. 
The first line in (\ref{A^S}) corresponds to the $s$-channel diagram 
exchanging the neutral gauge-scalar mode (\ie, the Higgs boson) 
while the second line represents 
the contributions from the $t$-channel and the $u$-channel exchanges 
of the charged gauge-scalar modes. 
Notice that the Higgs boson is massless at tree-level 
in the gauge-Higgs unification. 
It acquires a nonzero mass at one-loop order. 
 
The gauge invariance of the theory ensures the equivalence theorem~\cite{ET}, 
which states that the scattering of the longitudinally polarized vector bosons 
is equivalent to that of the (would-be) NG bosons eaten by the gauge bosons. 
In 5D models, the gauge-scalar modes~$\vph^{(n)}$ coming from $A_y$ play 
the role of the NG bosons in the equivalence 
theorem~\cite{Higgsless,5D_ET}.\footnote{
When there exist additional bulk Higgs fields that break the gauge symmetry, 
the NG bosons become mixture of modes from the bulk Higgs fields and $A_y$. }
Namely, the following relation holds for the longitudinal vector 
modes~$A_L^{(n)}$. 
\be
 T(A_L^{(n_1)},\cdots,A_L^{(n_l)};\Phi) 
 = C_l T(i\vph^{(n_1)},\cdots,i\vph^{(n_l)};\Phi)+\cO\brkt{\frac{M^2}{E^2}}, 
 \label{Eq_Th}
\ee
where all external lines are directed inwards, 
$\Phi$ denotes any possible amputated external physical fields, 
such as the transverse gauge boson, and $M$ is the heaviest mass 
among the external lines.  
A constant~$C_l$ is gauge-dependent, but $C_l=1$ at tree-level.\footnote{
We can also take a gauge where $C_l=1$ at all orders of the perturbative 
expansion~\cite{ET_loop}. 
}
The correction term is $\cO(M^2/E^2)$ 
because of the 5D gauge invariance (see Ref.~\cite{tanabashi}, for example). 
Eq.(\ref{Eq_Th}) is useful to discuss the high-energy behavior of 
the scattering amplitude~$\cA$ because the corresponding NG boson amplitude 
does not have $\cO(E^4)$ contributions,\footnote{
For the non-forward (non-backward) scattering, $\cO(E^2)$ contributions 
are also absent. } 
which makes it easier to numerically calculate the amplitude 
thanks to the absence of cancellations between large numbers. 

The scattering amplitude for the corresponding NG bosons 
comes only from the diagrams exchanging the vector modes. 
\bea
 \cB \eql ig_5^2\sum_{\alp,\bt=1,4}\int_0^L\dr y\int_0^L\dr y'\;
 V_{WZ}^\alp(y)(p_1+p_3)^\mu G_{\mu\nu}^{\alp\bt}(p_{13},y,y')
 (p_2+p_4)^\nu V_{WZ}^\bt(y') \nonumber\\
 &&+ig_5^2\sum_{\alp,\bt=1,4}\int_0^L\dr y\int_0^L\dr y'\;
 V_{WZ}^\alp(y)(p_1+p_4)^\mu G_{\mu\nu}^{\alp\bt}(p_{14},y,y')
 (p_2+p_3)^\nu V_{WZ}^\bt(y'), 
 \label{expr:cB}
\eea
where
\bea
 V_{WZ}^1 \defa \frac{e^{-2\sgm}}{2}
 \brkt{v_W^1v_Z^{3'}+v_W^4v_Z^6}, \nonumber\\
 V_{WZ}^4 \defa \frac{e^{-2\sgm}}{2}\brkt{v_W^1z_Z^6-v_W^4v_Z^{3'}}. 
\eea

\subsection{Various behaviors of the amplitudes} \label{behaviors}
Here we show various behaviors of the scattering amplitudes given 
in the previous subsection. 
For the numerical calculation, we choose the gauge parameter as $\xi=1$, 
the 4D gauge coupling~$g_4\equiv g_5/\sqrt{L}$ as $g_4^2=0.1$, 
and take the $W$ boson mass~$m_W$ as an input parameter. 
Then the size of the extra dimension~$L$ becomes $\tht$-dependent 
after fixing $m_W$. 
(See Eq.(\ref{mWZ:flat}), for example.)
The KK mass scale~$m_{\rm KK}\equiv \pi k/(e^{kL}-1)$ is also 
$\tht$-dependent. 
Thus the amplitudes are functions of the center-of-mass energy~$E$, 
the Wilson line phase~$\tht$ and the warp factor~$e^{kL}$.
The physical amplitude~$\cA$ is of course gauge-independent, 
and the $\xi$-dependence of the NG boson amplitude~$\cB$ is small 
in high-energy region as can be seen from (\ref{Eq_Th}). 
The 4 momenta and the polarization vectors of the initial and final states 
are parameterized as in Table.~\ref{Ep}. 
There, $p_W\equiv\sqrt{E^2/4-m_W^2}$, $p_Z\equiv\sqrt{E^2/4-m_Z^2}$, 
and $\phi$ is the scattering angle in the center-of-mass frame. 
Notice that the amplitudes~$\cA$ in (\ref{expr:cA}) and 
$\cB$ in (\ref{expr:cB}) are symmetric under $\phi\exch\pi-\phi$. 
\begin{table}[t]
\begin{center}
\begin{tabular}{|l|l|} \hline
$p_1=(E/2,0,0,p_W)$ & $\vep_1(p_1)=(p_W,0,0,E/2)/m_W$ \\
$p_2=(E/2,0,0,-p_W)$ & $\vep_2(p_2)=(p_W,0,0,-E/2)/m_W$ \\
$p_3=(E/2,p_Z\sin\phi,0,p_Z\cos\phi)$ & 
$\vep_3(p_3)=(p_Z,(E/2)\sin\phi,0,(E/2)\cos\phi)/m_Z$ \\
$p_4=(E/2,-p_Z\sin\phi,0,-p_Z\cos\phi)$ & 
$\vep_4(p_4)=(p_Z,-(E/2)\sin\phi,0,-(E/2)\cos\phi)/m_Z$ \\ \hline
\end{tabular}
\end{center}
\caption{The 4 momenta and the polarization vectors of the initial 
and the final states for $W_L^+(p_1)+W_L^-(p_2)\to Z_L(p_3)+Z_L(p_4)$. 
$E$ is the center-of-mass energy,  
$p_W\equiv\sqrt{E^2/4-m_W^2}$, $p_Z\equiv\sqrt{E^2/4-m_Z^2}$, 
and $\phi$ is the scattering angle in the center-of-mass frame. }
\label{Ep}
\end{table}

\subsubsection{Non-forward scattering}
\begin{figure}[t]
\centering  \leavevmode
\includegraphics[width=70mm]{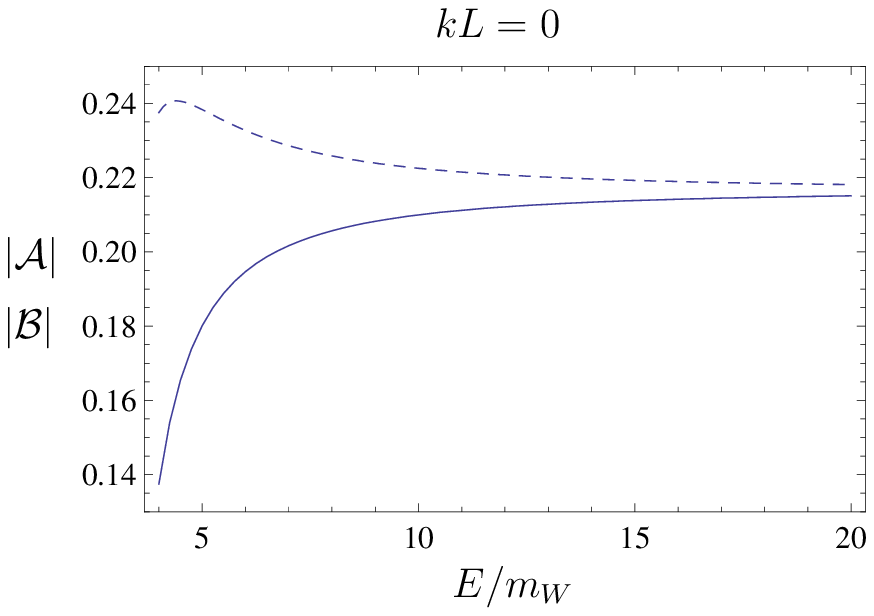} \hspace{10mm}
\includegraphics[width=70mm]{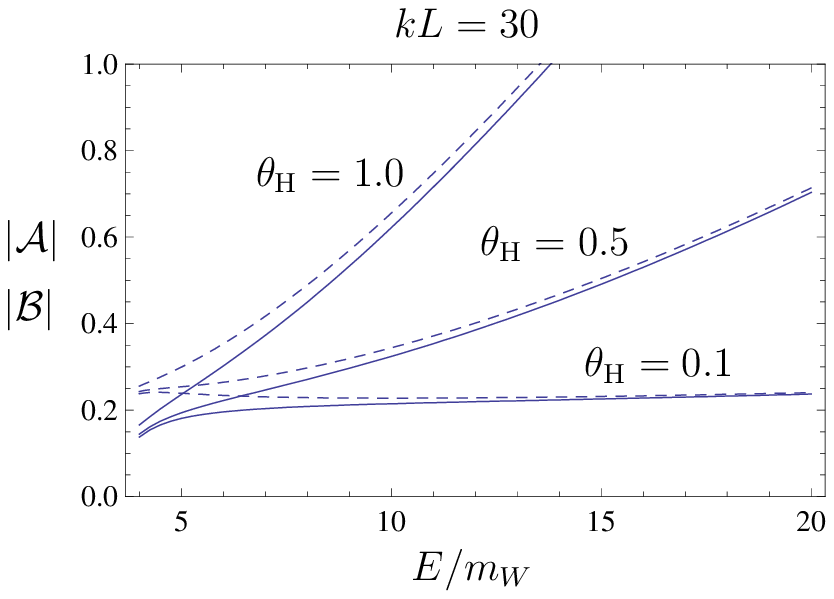}
\caption{The energy dependence of the amplitudes 
for $W^+_L+W^-_L\to Z_L+Z_L$. 
The solid lines represent the vector boson scattering~$\cA$, 
and the dashed lines are the NG boson scattering~$\cB$. 
The scattering angle is chosen as $\phi=\pi/3$. 
In the flat case (the left figure), the amplitudes are 
independent of the Wilson line phase~$\tht$ for $0<\tht<\pi/2$. }
\label{WWZZ:cA-E}
\end{figure}
First we consider a non-forward (and non-backward) scattering. 
We choose the scattering angle as $\phi=\pi/3$ in the following. 
Fig.~\ref{WWZZ:cA-E} shows the energy dependence of 
the scattering amplitudes. 
The solid and the dashed lines represent 
the scattering amplitudes for the vector bosons~$\cA$ and 
for the NG bosons~$\cB$, respectively. 
We can explicitly see that the equivalence theorem holds both in the flat 
and the warped cases, and $\abs{\cB}-\abs{\cA}=\cO(m_W^2/E^2)$. 

In the flat case ($kL=0$), the amplitude~$\cA$ is 
independent of the Wilson line phase~$\tht$ in the range of 
$0<\tht<\pi/2$. 
It approaches to a constant value at high energies. 
In the case of the warped geometry, on the other hand, 
The amplitude has a large $\tht$-dependence and increase as $E^2$. 
It grows faster for larger $\tht$. 

\begin{figure}[t]
\centering  \leavevmode
\includegraphics[width=70mm]{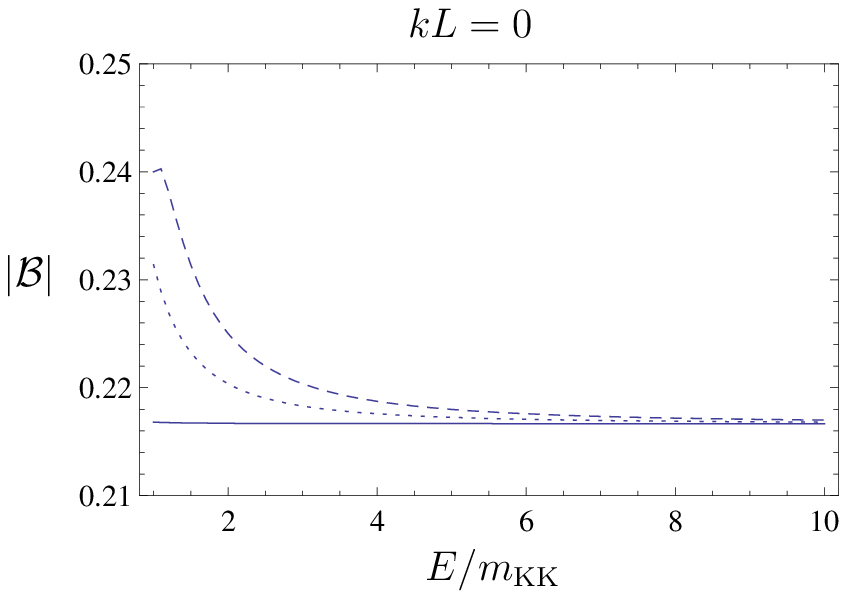} \hspace{10mm}
\includegraphics[width=70mm]{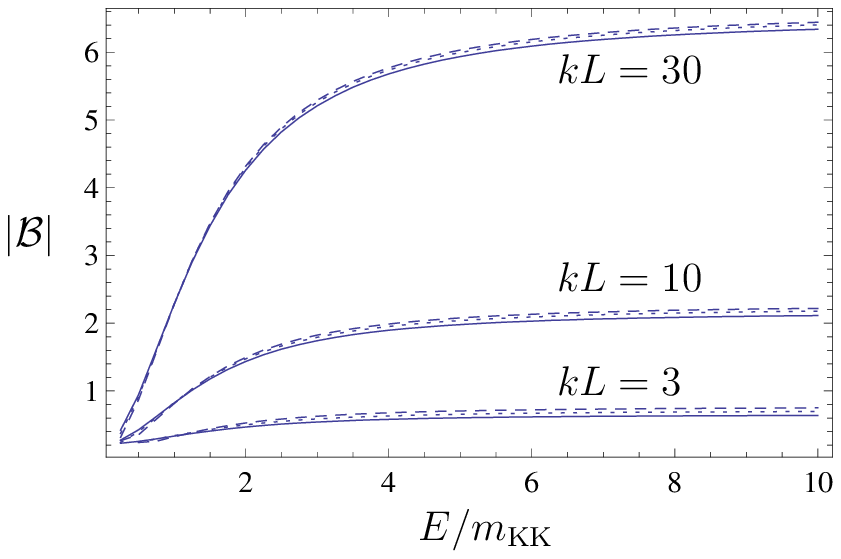}
\caption{The energy dependence of the amplitude 
in the unit of $m_{\rm KK}$. 
The solid, dotted, dashed lines correspond to $\tht=0.1,1.0,1.5$, 
respectively. 
}
\label{WWZZ:cA-E2}
\end{figure}

These behaviors reflect the $\tht$-dependences of the coupling constants 
among the gauge and the Higgs modes and of the KK mass scale~$m_{\rm KK}$. 
Before explaining the behaviors of the amplitude, let us see 
the energy dependence of $\cB$ again 
by rescaling the unit of the horizontal axes 
to $m_{\rm KK}$. (Fig.~\ref{WWZZ:cA-E2})
Then we can see that the amplitude approaches to constant values 
at sufficiently high energies even in the warped case. 
The constant values vary depending on the warp factor, 
and are larger than the value in the flat case
by a factor~$kL$ for $kL\simgt\cO(1)$. 
The $\tht$-dependence we have seen in the right plot of Fig.~\ref{WWZZ:cA-E} 
now almost disappears in the unit of $m_{\rm KK}$. 
It is cancelled by the $\tht$-dependence of $m_{\rm KK}$
(see the beginning of Sec.~\ref{behaviors}.).
The apparent $\tht$-dependence of the plots in the flat case stems from 
the $\tht$-dependence of $m_{\rm KK}$. 

Now we will interpret the above behaviors of the amplitude. 
First of all, we should notice that the model reduces to 
the ``standard model'' (SM), in which the Weinberg angle is 
$\sin^2\theta_W=3/4$ and the Higgs boson 
is massless, when $\tht\ll 1$ irrespective of the 5D geometry. 
Every coupling constant in the gauge-Higgs sector 
takes almost the SM value and the KK modes 
are heavy enough to decouple. 
Thus the amplitude takes the same value as SM 
up to the energy scale where the KK modes start to propagate, \ie, 
$m_{\rm KK}$. 
The amplitude takes an almost constant value at 
$E\simgt\cO(10m_W)$ in this case. 


When $\tht$ is not small, the coupling constants relevant to the amplitude 
deviate from the SM value. 
In the warped case, the $WWH$ and the $ZZH$ couplings become smaller 
than the SM values by $\cos(\tht/2)$ and $\cos\tht$ respectively, 
while the $WWZZ$ and the $WWZ$ couplings are almost 
unchanged~\cite{GHU:HS1,GHU:HS2}. 
Thus $\cO(E^2)$ contributions miss to be cancelled among the low-lying modes 
and the amplitude grows in the low-energy region. 
For larger value of $\tht$ (up to $\pi/2$), the deviation of the couplings 
are larger and then the amplitude grows faster. 
(See the right figure of Fig.~\ref{WWZZ:cA-E}.)
This remaining $\cO(E^2)$ contribution is eventually cancelled by 
contributions from the KK modes. 
Namely, the amplitude ceases to increase and approaches to a constant value 
when the KK modes start to propagate. 

The flat spacetime is a special case. 
As we mentioned, the amplitude becomes almost constant 
at $E\simgt\cO(10m_W)$ when $\tht\ll 1$. 
For larger values of $\tht$, 
the $WWZZ$ and the $WWZ$ couplings slightly deviate from 
the SM values because of the nontrivial $y$-dependences 
of the mode functions for the $W$ and the $Z$ bosons~\cite{GHU:HS2}, 
while the $WWH$ and the $ZZH$ couplings are now unchanged. 
Then the $\cO(E^2)$ contributions fail to be cancelled 
among the low-lying modes, just like in the warped case. 
However 
the contribution from the KK-modes {\it completely} cancel 
this $\cO(E^2)$,  
and the amplitude results in unchanged from the $\tht\ll 1$ case. 
Namely the effect of the $\tht$-dependence of the $WWZZ$ and $WWZ$ couplings 
and that of the KK mass spectrum are completely cancelled and 
the amplitude becomes $\tht$-independent for $0<\tht\leq\pi/2$ 
in the flat case. 
In the range of $\pi/2<\tht<\pi$, the amplitude has a nontrivial 
$\tht$-dependence. 
This stems from the fact that the relation~$m_Z/m_W=2$ no longer holds 
(see Eq.(\ref{mWZ:flat}))
and $m_Z$ also has a nontrivial $\tht$-dependence in this region.

\subsubsection{Forward scattering}
\begin{figure}[t]
\centering  \leavevmode
\includegraphics[width=70mm]{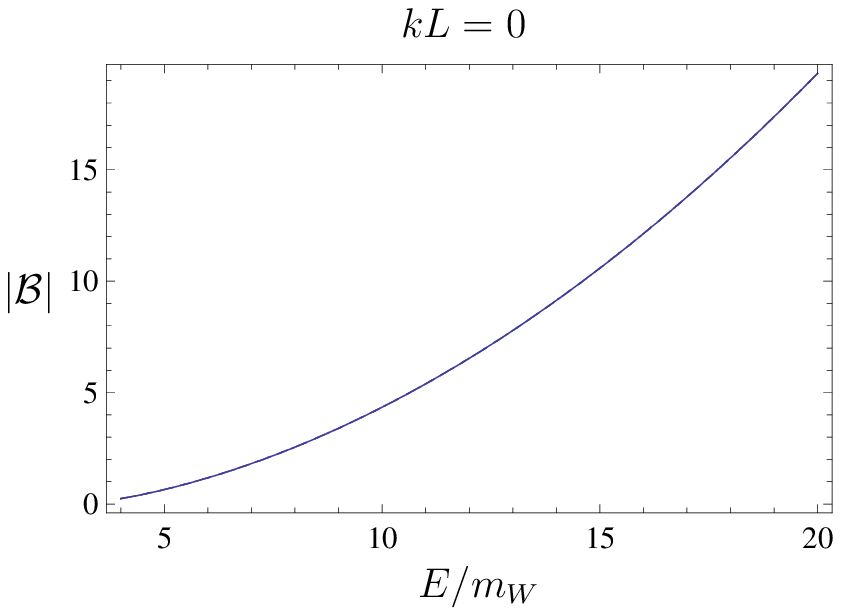} \hspace{10mm}
\includegraphics[width=70mm]{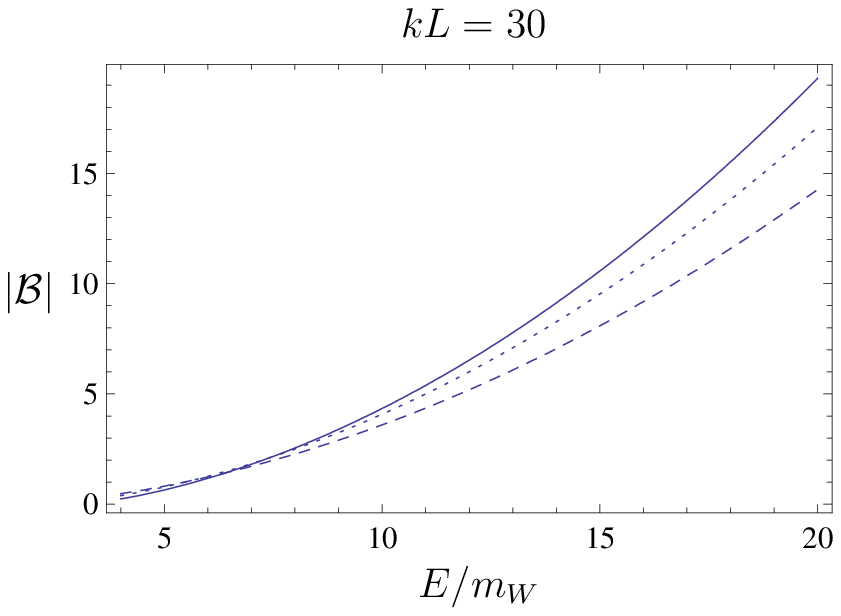}
\caption{The energy dependence of the amplitude at $\phi=0$. 
The solid, dotted, dashed lines correspond to $\tht=0.1,1.0,1.5$, 
respectively. 
}
\label{WWZZ:cA-E3}
\end{figure}

Next we consider the forward scattering, \ie, $\phi=0$. 
In this case, an $\cO(E^2)$ contribution remains 
and the amplitude monotonically increases even above $m_{\rm KK}$. 
This is because the power counting of $E$ for the amplitude changes 
around $\phi=0$. 
For example, the brace part in (\ref{A^S}) 
is expanded (for nonzero $\sin\phi$) as 
\bea
 A_{tu} \defa 
 \frac{(\vep_1\cdot\vep_3^*)(\vep_2\cdot\vep_4^*)}{p_{13}^2}
 +\frac{(\vep_1\cdot\vep_4^*)(\vep_2\cdot\vep_3^*)}{p_{14}^2} \nonumber\\
 \eql \frac{E^2}{4m_W^2m_Z^2}-\frac{m_W^2+m_Z^2}{2m_W^2m_Z^2} 
 +\frac{2m_W^2m_Z^2+(m_W^4+m_Z^4)\cos(2\phi)}{m_W^2m_Z^2E^2\sin^2\phi}
 +\cO(E^{-4}). 
\eea
This means that the expansion becomes invalid when $\sin\phi\simlt\cO(m_W/E)$. 
At $\phi=0$, this quantity reduces to 
\bea
 A_{tu} \eql \frac{(m_W^4+m_Z^4)E^2}{2m_W^2m_Z^2(m_Z^2-m_W^2)^2}
 -\frac{2(m_W^2+m_Z^2)}{(m_Z^2-m_W^2)^2}, 
\eea
and the leading term for the high energy expansion changes. 
Therefore an $\cO(E^2)$ contribution is left in the total amplitude. 
Similar behavior of the amplitude is observed also in the standard model. 
Fig.~\ref{WWZZ:cA-E3} shows the energy dependence of 
the forward scattering amplitude. 
We can see that the amplitude grows as $E^2$ in any cases. 
In the flat case, the amplitude does not have the $\tht$-dependence again. 
In the warped case, it varies for different values 
of $\tht$. 
For small values of $\tht$, the amplitude has little dependence on 
the warp factor and takes almost the same value as the flat case. 
For larger values of $\tht$, it becomes smaller in contrast to 
the non-forward scattering.

\subsubsection{S-wave amplitude}
\begin{figure}[t]
\centering  \leavevmode
\includegraphics[width=70mm]{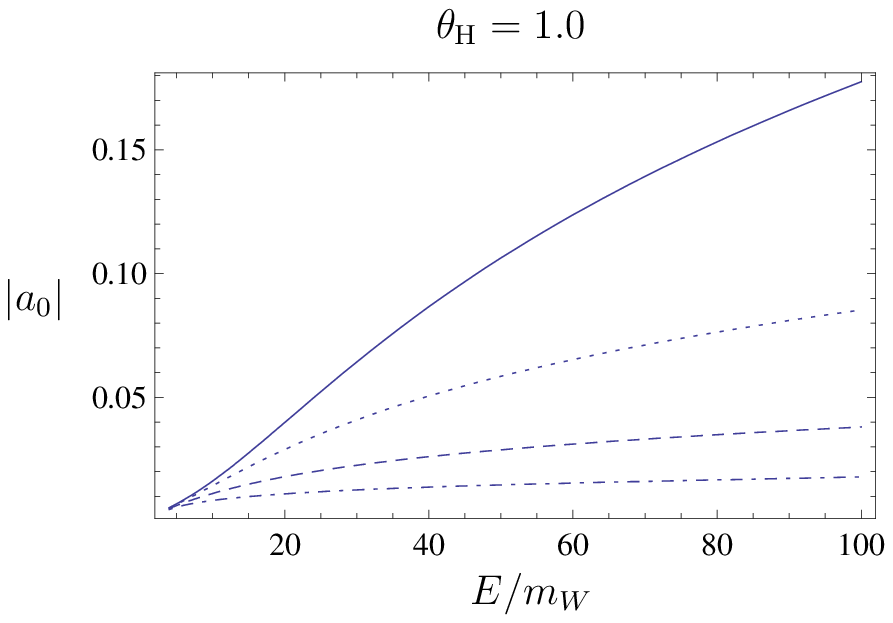} \hspace{10mm}
\includegraphics[width=70mm]{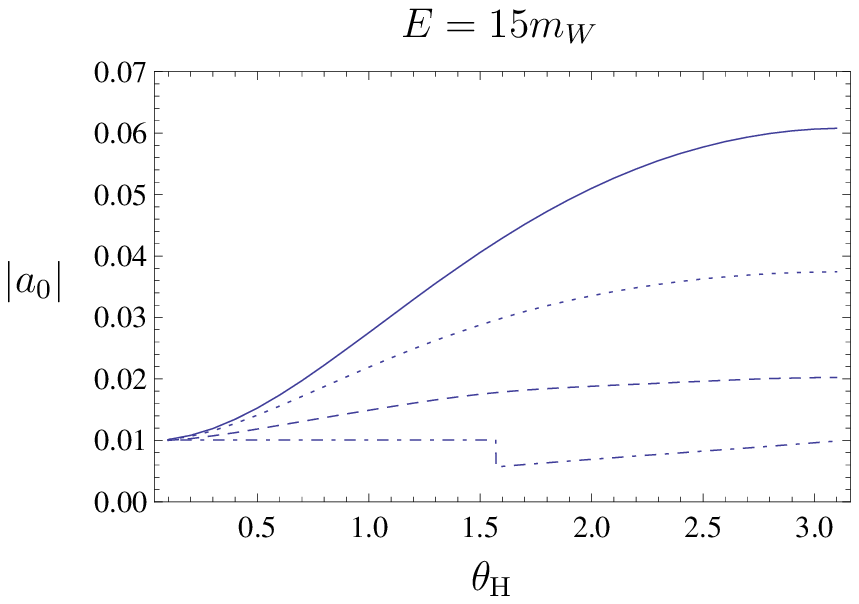}
\caption{The s-wave amplitude as functions of $E$ and $\tht$. 
The solid, dotted, dashed, dotdashed lines correspond to 
$kL=30,10,3,0$, respectively. 
}
\label{S-wave}
\end{figure}

The conventional bound for the tree-level unitarity is given by\footnote{
More restrictive unitarity condition is proposed in Ref.~\cite{DJL}. 
} 
\be
 \abs{a_0}\leq 1, 
\ee
where $a_0$ is the s-wave amplitude defined as 
\bea
 a_0(E) \defa \frac{1}{32\pi}\int_{-1}^1 \dr (\cos\phi)\;\cA(E,\cos\phi). 
 \label{def:a_0}
\eea
Hence we now estimate the s-wave amplitude. 
As we mentioned above, the integrand grows as $E^2$ 
in the region~$1-\abs{\cos\phi}\simlt\cO(m_W^2/E^2)$
while it approaches to a constant for large $E$ 
in the other region of $\cos\phi$. 
Therefore $a_0(E)$ behaves as $\cO(E^0)$ at high energies. 
In fact, it grows logarithmically in high-enery region as shown in 
the left plot of Fig.~\ref{S-wave}. 
The right plot shows the $\tht$-dependence of 
$a_0(15m_W)$. 
We can see from these plots that the amplitude becomes larger 
for larger warp factor and larger $\sin(\tht/2)$.\footnote{
Notice that the period of $\tht$ is $2\pi$. 
(See the first equation in Eq.(\ref{eq_for_spectrum}).)} 
In the flat limit, it decreases and has only a small $\tht$-dependence. 
In fact, it is independent of $\tht$ for $0<\tht\leq \pi/2$. 
The small $\tht$-dependence for $\pi/2<\tht<\pi$ originates from 
the fact that the relation~$m_Z/m_W=2$ no longer holds 
and $m_Z$ becomes $\tht$-dependent, which is peculiar to 
the $SU(3)$ model.  
Thus the details of the small $\tht$-dependence in the flat case 
is model-dependent. 

In the above calculation, we have chosen the value of the gauge coupling as 
$g_4^2=0.1$. 
Since the tree-level amplitude is proportional to $g_4^2$, 
it becomes four times larger if we take $g_4$ as the weak gauge coupling 
in the standard model. 
In this case, a scale~$\Lmd$ determined by $a_0(\Lmd)=1$ 
is estimated as $\Lmd\sim 1500m_W$ for $\tht=0.1$ and 
$\Lmd\sim 150m_W$ for $\tht=1.5$ when $kL=30$, for example.  
This suggests that the perturbative unitarity will be violated at a lower scale 
for larger $\sin(\tht/2)$. 
In order to estimate the unitarity bound, we have to consider 
other scattering processes and sum up all the possible final states 
including the KK states. 
Thus the real cut-off scale~$\Lmd_{\rm cut}$ is expected as much lower scale 
than the above values of $\Lmd$. 
In particular, in the latter example where $m_{\rm KK}\simeq 18m_W$ 
and the Higgs mode hardly contributes to the unitarization, 
it is expected that $\Lmd_{\rm cut}$ becomes around 1~TeV 
as in the standard model without the Higgs field.

\section{Summary and discussions} \label{summary}
We have investigated the weak boson scattering in the gauge-Higgs unification, 
focusing on the dependence of the amplitude on the scattering energy~$E$, 
the Wilson line phase~$\tht$ and the warp factor~$e^{kL}$. 
In this paper we consider a process: $W^+_L+W^-_L\to Z_L+Z_L$ 
in the $SU(3)$ model as the simplest example. 

The 5D propagators are useful to calculate the scattering amplitudes 
because we need not explicitly calculate the KK mass spectra nor 
perform the infinite summation over the KK modes propagating 
in the internal lines. 
We have numerically checked the equivalence theorem between 
the amplitudes for the longitudinal vector bosons and 
the (would-be) NG bosons. 
The correction term is read off as $\cO(m_W^2/E^2)$. 

The amplitude behaves differently in the flat and the warped spacetimes. 
It is independent of $\tht$ in the flat case, 
while a nontrivial $\tht$-dependence comes out in the warped case. 
These behaviors come from the $\tht$-dependences of the coupling constants 
among the gauge and the Higgs modes and of the KK mass scale~$m_{\rm KK}$ 
in the case that the $W$ boson mass~$m_W$ is fixed as an input parameter  
(see the beginning of Sec.~\ref{behaviors}.).
For the non-forward (and non-backward) scattering, 
the amplitude approaches to a constant at high energies in both cases, 
but the asymptotic constant value is enhanced by a factor~$kL$ 
in the warped case ($kL\simgt\cO(1)$), 
comparing to that in the flat case. 
On the other hand, the forward (backward) scattering amplitude grows as $E^2$. 
The s-wave amplitude grows logarithmically in high energy region 
just like in the standard model, 
and depends on $\tht$ in the warped case. 
Thus, even if we consider only the process~$W^+_L+W^-_L\to Z_L+Z_L$, 
the tree-level unitarity will be violated for quite large $E$.
It is known, however, in higher dimensional theories,
the unitarity violation appears at a lower energy, by
summing up all the possible final states, exhibiting
the non-renormalizability. 
Generically the amplitude is enhanced in the warped case for $\tht=\cO(1)$. 
This suggests that the tree-level unitarity will be violated 
at a lower scale in the warped case than the flat case. 

In Ref.~\cite{FPR}, three separate scales that determine the dynamics 
of the scattering process are introduced, \ie, 
the electroweak breaking scale~$v$, 
the Higgs boson decay constant~$f_h$,\footnote{
This is the composite scale of the Higgs boson in the holographic dual picture. 
}
and the KK scale~$\mKK$. 
In our notation, these scales are related to each other as 
$v=f_h\tht/2$ and $f_h=\sqrt{2}/(g_5\sqrt{L})=\sqrt{2}\mKK/(\pi g_4)$ 
in the flat case, and 
$v=f_h\sin(\tht/2)$ and $f_h\simeq 2\sqrt{k}e^{-kL}/g_5 
\simeq 2\mKK/(\pi g_4\sqrt{kL})$ in the warped case. 
In the terminology of Ref.~\cite{FPR}, the case of $\tht\ll 1$ is 
referred to as the `Higgs limit', and the case of $\tht=\cO(1)$ is 
as the `Higgsless limit'. 
The Higgs boson unitarizes the scattering process in the former 
while it does not (or does only partly) in the latter. 

For the purpose of estimating the scale of the unitarity violation, 
we should extend our analysis for the following points. 
We should take into account the Higgs mass, 
which is induced by the quantum effect, 
and the decay widths of the weak bosons. 
The latter is necessary to discuss the process: $W^+_L+W^-_L\to W^+_L+W^-_L$, 
for example. 
The infrared singularity for the forward scattering of this process 
is smeared out by taking into account the width of the $W$ boson. 
Furthermore, we have to sum up all the possible final states 
including the KK states to discuss the unitarity. 
Since the $SU(3)$ model is a toy model, we should work in a more realistic 
model, for example, the $SO(5)\times U(1)$ model~\cite{Wagner,GHU:HS1,GHU:HS2}. 
In the flat spacetime, the spectrum of the latter model has a 
qualitatively different $\tht$-dependence from the former 
due to the nontrivial boundary conditions of the 5D gauge fields.\footnote{
These boundary conditions are effectively obtained from the orbifold ones 
by introducing some boundary terms. }
Each mass eigenvalue is not a linear function of $\tht$ 
(see Fig.~1 in Ref.~\cite{GHU:HS2}) in contrast to the $SU(3)$ model. 
This difference may affects the $\tht$-independence 
of the scattering amplitude found in the our model. 
These issues will be discussed in a subsequent paper.

\vskip 0.5cm

\leftline{\bf Acknowledgments} \nopagebreak
The authors would like to thank Y. Hosotani, M.~Tanabashi 
and K.~Tobe for useful discussions and comments. 
This work was supported in part 
by the Japan Society for the Promotion of Science (T.Y.) and 
by Special Postdoctoral Researchers Program at RIKEN (Y.S.).

\appendix

\section{Bases of mode functions} \label{basis_fcn}
Here we define bases of mode functions, 
following Ref.~\cite{Falkowski}. 
The functions~$C_0(y,m)$ and $S_0(y,m)$ are defined as 
two independent solutions to 
\be
 \brkt{\frac{d}{dy}e^{-2\sgm}\frac{d}{dy}+m^2}f=0, 
\ee
with initial conditions 
\bea
 C_0(0,m) \eql 1, \;\;\;
 C'_0(0,m) = 0, \nonumber\\
 S_0(0,m) \eql 0, \;\;\;
 S'_0(0,m) = me^{-\sgm(L)}.  \label{basis_fcn1}
\eea

For the derivation of 5D propagators in Appendix~\ref{5Dpropagator}, 
it is convenient to define another basis functions~$C_L(y,m)$ 
and $S_L(y,m)$ with initial conditions 
\bea
 C_L(L,m) \eql 1, \;\;\;
 C'_L(L,m) = 0, \nonumber\\
 S_L(L,m) \eql 0, \;\;\;
 S'_L(L,m) = me^{\sgm(L)}. \label{basis_fcn2}
\eea

From the Wronskian relation, the above functions satisfy 
\bea
 && C_0(y,m)S'_0(y,m)-S_0(y,m)C'_0(y,m) \nonumber\\
 \eql C_L(y,m)S'_L(y,m)-S_L(y,m)C'_L(y,m)
 = me^{2\sgm(y)-\sgm(L)}.  \label{Wronskian}
\eea

The two bases are related to each other by 
\bea
 C_L(y,m) \eql \frac{e^{-\sgm(L)}}{m}\brc{S'_0(L,m)C_0(y,m)
 -C'_0(L,m)S_0(y,m)}, \nonumber\\
 S_L(y,m) \eql -\brc{
 S_0(L,m)C_0(y,m)-C_0(L,m)S_0(y,m)}. 
 \label{rel:basis_fcns}
\eea

\begin{description}
\item[Flat spacetime] \mbox{}\\
In the flat spacetime, \ie, $\sgm(y)=0$, 
the basis functions are reduced to 
\bea 
 C_0(y,m) \eql \cos(my), \;\;\;
 S_0(y,m) = \sin(my), \nonumber\\
 C_L(y,m) \eql \cos\brc{m(y-L)}, \;\;\;
 S_L(y,m) = \sin\brc{m(y-L)}. 
\eea

\item[Randall-Sundrum spacetime] \mbox{}\\
In the Randall-Sundrum spacetime, \ie, $\sgm(y)=ky$, 
the basis functions are written in terms of the Bessel functions as 
\bea
 C_0(y,m) \eql \frac{\pi m}{2k}e^{ky}
 \brc{Y_0\brkt{\frac{m}{k}}J_1\brkt{\frac{m}{k}e^{ky}}
 -J_0\brkt{\frac{m}{k}}Y_1\brkt{\frac{m}{k}e^{ky}}}, \nonumber\\
 S_0(y,m) \eql -\frac{\pi m}{2k}e^{k(y-L)}
 \brc{Y_1\brkt{\frac{m}{k}}J_1\brkt{\frac{m}{k}e^{ky}}
 -J_1\brkt{\frac{m}{k}}Y_1\brkt{\frac{m}{k}e^{ky}}}, \nonumber\\
 C_L(y,m) 
 \eql \frac{\pi m}{2k}e^{ky}
 \brc{Y_0\brkt{\frac{m}{k}e^{kL}}J_1\brkt{\frac{m}{k}e^{ky}}
 -J_0\brkt{\frac{m}{k}e^{kL}}Y_1\brkt{\frac{m}{k}e^{ky}}}, \nonumber\\
 S_L(y,m) 
 \eql -\frac{\pi m}{2k}e^{ky}
 \brc{Y_1\brkt{\frac{m}{k}e^{kL}}J_1\brkt{\frac{m}{k}e^{ky}}
 -J_1\brkt{\frac{m}{k}e^{kL}}Y_1\brkt{\frac{m}{k}e^{ky}}}. \nonumber\\
\eea
\end{description}

\section{Derivation of 5D propagators} \label{5Dpropagator}
Here we derive explicit forms of 5D propagators. 
We take the same strategy as in the appendix of Ref.~\cite{GP}. 
Since the 4D vector part~$A_\mu$ and the gauge-scalar part~$A_y$ are decoupled 
at the quadratic level with our choice of the gauge-fixing function, 
the mixed components of 
the propagator~$\langle 0|TA_\mu^\alp(p,y)A_y^\bt(-p,y')|0\rangle$ vanish. 
In this section, we work in the Scherk-Schwarz basis defined 
by (\ref{gauge_trf}) and (\ref{def:Omg}). 

\subsection{Vector propagator}
The gauge index~$\alp=1,\cdots,8$ is decomposed into two parts as 
$a=1,2,3',8'$ and $\hat{a}=4,5,6,7$, according to the $Z_2$-parities 
of $A_\mu^\alp$. 
Then the 5D propagator~$iG_{\mu\nu}^{\alp\bt}(p,y,y')\equiv
\langle 0|TA_\mu^\alp(p,y)A_\nu^\bt(-p,y')|0\rangle$ satisfies 
\be
 \sbk{\brc{\der_y^2-2\sgm'\der_y-e^{2\sgm}p^2}
 \dlt_\mu^{\;\;\nu}+e^{2\sgm}\brkt{\frac{1}{\xi}-1}
 p_\mu p^\nu}G^{\alp\bt}_{\nu\rho}(p,y,y') 
 = e^{2\sgm}\eta_{\mu\rho}\dlt^{\alp\bt}\dlt(y-y'),  
 \label{eq_propagator}
\ee
with the boundary conditions, 
\bea
 &&\der_y G_{\mu\nu}^{a\bt}(p,0,y') = G_{\mu\nu}^{\hat{a}\bt}(p,0,y') = 0, 
 \nonumber\\
 &&\brkt{R_\theta}^{a\gm}\der_yG_{\mu\nu}^{\gm\bt}(p,L,y') 
 = \brkt{R_\theta}^{\hat{a}\gm}G_{\mu\nu}^{\gm\bt}(p,L,y') = 0, 
 \label{bdcd_propagator}
\eea
where a constant matrix~$R_\theta$ is a rotation matrix for the indices 
of the adjoint representation corresponding to a transformation by 
$\Omg(L)$ defined in (\ref{def:Omg}), \ie, 
\be
 \brkt{R_\theta}^{\alp\bt}A_M^\bt = \sbk{\Omg^{-1}(L)A_M\Omg(L)}^\alp 
 \equiv \tr\brc{\lmd^\alp\Omg^{-1}(L)A_M\Omg(L)}. 
 \label{def:R_tht}
\ee

We can decompose $G_{\mu\nu}^{\alp\bt}(p,y,y')$ into 
the following two parts. 
\be
 G_{\mu\nu}^{\alp\bt}(p,y,y')=\brkt{\eta_{\mu\nu}-\frac{p_\mu p_\nu}{p^2}}
 G_{\rm T}^{\alp\bt}(y,y',\abs{p})
 +\frac{p_\mu p_\nu}{p^2}G_{\rm S}^{\alp\bt}(y,y',\abs{p}),  
\ee 
where $\abs{p}\equiv\sqrt{-p^2}$. 
The first and the second terms correspond to the propagators for 
$A_\mu^{(n)}$ and $A_{\rm S}^{(n)}$, respectively. 
Writing $G_{\rm T}^{\alp\bt}(y,y',\abs{p})$ as 
\be
 G_{\rm T}^{\alp\bt}(y,y',\abs{p})=\vth(y-y')G_{\rm T>}^{\alp\bt}(y,y',\abs{p})
 +\vth(y'-y)G_{\rm T<}^{\alp\bt}(y,y',\abs{p}), 
\ee
the solutions to (\ref{eq_propagator}) satisfying 
(\ref{bdcd_propagator}) are given in the matrix notation 
for the index~$\alp=(a,\hat{a})$ by 
\bea
 G_{\rm T<}(y,y',\abs{p}) \eql \cM_0(y,\abs{p})\alp_{\rm T<}(y',\abs{p}), 
 \nonumber\\
 R_\theta G_{\rm T>}(y,y',\abs{p}) \eql \cM_L(y,\abs{p})\alp_{\rm T>}(y',\abs{p}),  
 \label{expr:Gp1}
\eea
where 
\be
 \cM_0 \equiv \begin{pmatrix} C_0\cdot\id_4 & \\
 & S_0\cdot\id_4 \end{pmatrix}, \;\;\;
 \cM_L \equiv \begin{pmatrix} C_L\cdot\id_4 & \\ 
 & S_L\cdot\id_4 \end{pmatrix}. 
 \label{def:cM}
\ee
The unknown matrix functions~$\alp_{\rm T<}(y',\abs{p})$ and 
$\alp_{\rm T>}(y',\abs{p})$ are determined 
by imposing the following matching conditions at $y=y'$. 
The continuity of $G_{\rm T}$ at $y=y'$ leads to the condition
\be
 G_{\rm T<}(y,y,\abs{p}) = G_{\rm T>}(y,y,\abs{p}), 
 \label{continuity}
\ee
and we obtain from (\ref{eq_propagator}) the condition
\be
 \left\{\der_yG_{\rm T>}(y,y',\abs{p})
 -\der_yG_{\rm T<}(y,y',\abs{p})\right\}_{y'\to y}
 = e^{2\sgm(y)}.  \label{discontinuity}
\ee
Using these conditions, we obtain the 5D propagators as 
\bea
 G_{\rm T<}(y,y',\abs{p}) \eql 
 e^{2\sgm(L)}\cM_0(y,\abs{p})\cW^{-1}(\abs{p})\cM_L(y',\abs{p})R_\theta, 
 \nonumber\\
 G_{\rm T>}(y,y',\abs{p}) \eql \brc{G_{\rm T<}(y',y,\abs{p})}^t, 
 \label{expr:G_T}
\eea
where 
\bea
 \cW(\abs{p}) \defa e^{-2\sgm(y)+2\sgm(L)}
 \brkt{\cM'_L R_\theta\cM_0-\cM_L R_\theta\cM'_0}(y,\abs{p})  \label{def:cW}
\eea
is $y$-independent from the Wronskian relation~(\ref{Wronskian}). 

The part of the scalar modes~$G_{\rm S}(y,y',\abs{p})$ is obtained 
in a similar way, and it is related to $G_{\rm T}(y,y',\abs{p})$ as 
\be
 G_{\rm S}(y,y',\abs{p}) = G_{\rm T}(y,y',\abs{p}/\sqrt{\xi}). 
\ee

\subsection{Gauge-scalar propagator}
Next we consider the propagators for the gauge-scalar modes. 
The 5D propagator~$iG_{yy}^{\alp\bt}(y,y',\abs{p})\equiv
\langle 0|TA_y^\alp(p,y)A_y^\bt(-p,y')|0\rangle$ satisfies 
\be
  \brc{\xi\der_y^2 e^{-2\sgm}-p^2}
 G_{yy}^{\alp\bt}(y,y',\abs{p}) = e^{2\sgm}\dlt^{\alp\bt}\dlt(y-y'), 
 \label{eq_G:scalar}
\ee
with the boundary conditions, 
\bea
 G_{yy}^{a\bt}(0,y',\abs{p}) \eql 
 \der_y\brc{e^{-2\sgm}G_{yy}^{\hat{a}\bt}}(0,y',\abs{p}) = 0, 
 \label{bdcd0:scalar} \\
 \brkt{R_\theta}^{a\gm}G_{yy}^{\gm\bt}(L,y',\abs{p}) \eql 
 \der_y\brc{e^{-2\sgm}\brkt{R_\theta}^{\hat{a}\gm}
 G_{yy}^{\gm\bt}}(L,y',\abs{p}) = 0. 
 \label{bdcdL:scalar}
\eea
These can be solved by the same manner as in the previous subsection. 
We find that $G_{yy}(y,y',\abs{p})$ is related to 
$G_{\rm S}(y,y',\abs{p})$ as 
\bea
 G_{yy<}(y,y',\abs{p}) \eql 
 -\frac{1}{p^2}\der_y\der_{y'}G_{\rm S<}(y,y',\abs{p}),  \nonumber\\
 G_{yy>}(y,y',\abs{p}) \eql 
 -\frac{1}{p^2}\der_y\der_{y'}G_{\rm S>}(y,y',\abs{p}).  
 \label{rel:Gyy-Gs}
\eea


\end{document}